# Hyperspectral Spatial Super-Resolution using Keystone Error

Ankur Garg, Meenakshi Sarkar, S. Manthira Moorthi, Debajyoti Dhar

**ABSTRACT**
Hyperspectral Images aid in precise identification of objects on ground by measuring their signatures with very fine spectral resolution. If such images have high spatial resolution, it further boosts this ability. Increasing the spatial resolution with hardware by making use of bigger telescopes is a very costly affair in terms of the resources needed which makes the overall system optimization suffer. In comparison, increasing the spatial resolution with ground processing is the most optimal solution to handle this obstacle. Generally, to get advantage of both high spectral and spatial resolution data, hypersharpening approaches are used to merge high spectral resolution data with high spatial resolution. The approach gives good results when the two datasets have been acquired within very short time frame leading to similar atmospheric, viewing imaging conditions. This is usually possible when both high spectral and spatial resolution sensors are imaging from same platform. But using datasets from different dates and time often leads to radiometric artifacts which degrade the final image quality. In this work, we propose an approach to increase the spatial resolution of hyperspectral data without making use of any auxiliary input. The two key inputs for increasing the spatial resolution via superresolution algorithms are the high resolution point spread function and the sub-pixel shifts. For grating based hyperspectral sensors, misregistration between spectral channels occur due to instrument intrinsic keystone error which leads to sub-pixel shifts between spectral channels that vary across the spatial field of view. But the high resolution blur for such systems is seldom known due to the dynamic nature of the intervening atmosphere and platform jitter. Also, the optics blur for each channel differs due to differing wavelengths. To handle this dynamic effect, we first estimate the point spread function from lower resolution image using blind deconvolution and correct it using image deconvolution. Then, to handle the remaining sampling related blur, we assume model for it and use it in super resolution framework. This approach is different from earlier ones which assume that the high resolution blur present in the observation model is exactly known. The paper also proposes a new adaptive prior and shows how it performs better than those already present in literature. The algorithm is used to increase the resolution of visible and near infrared (VNIR) spectrometer of HySIS, ISRO's hyperspectral sensor, and the results shows that it is able to remove the aliasing and lead to a resolution boost of $\sim$1.3 times with respect to the input. The algorithm is generic and can be used to increase the resolution of similar configuration systems.

---

All authors are with Signal and Image Processing Group, Space Applications Center, Indian Space Research Organisation, Ahmedabad, India-380015, Corresponding email ids : (agarg, meenakshi, smmoorthi, deb)@sac.isro.gov.in

## 1.  Introduction

Spatial Resolution is the ability of an electro-optical sensor to discriminate the smallest object on the ground with specified accuracy. In a typical electro-optical sensor, spatial resolution is limited by factors such as diffraction limit of the optics, optical distortions like aberrations, detector sampling & averaging, the platform jitter, and intervening atmosphere between the object & sensor. Spectral Resolution of such sensors define the smallest spectral difference in which the signature of the target can be measured. It is limited by the signal to noise ratio of the imaging system which depends on the telescope size and the detector's full well capacity. There is always a trade-off between spatial and spectral resolution for any such system and various optimizations are done to select the optimal configuration. Hyperspectral sensors are a class of sensors with very high spectral resolution which defines their ability of see the target under observation with very fine spectral sampling. Because they are seeing the object under very small spectrum, to collect sufficient signal, generally, coarser spatial resolution is desired. This leads to decrease in the ability to discriminate between very small objects on ground. Hypersharpening (1) is a common approach which is used to increase the spatial resolution of sensor with higher spectral resolution (multispectral or hyperspectral) using some external sensor with higher spatial resolution (panchromatic or multispectral). The process works fine when both the sensors under usage are imaging from the same platform within very short time frame leading to similar atmospheric and viewing imaging conditions but they generally suffer from the errors occurring due to inter-sensor misregsitration between datasets. Also, when both such imaging sensors are not available on the same platform, data from external sources are used which further leads to radiometric and geometric distortions.

Hyperspectral systems are generally designed to have high telescope modulation transfer function at their Nyquist frequency which leads to aliasing, i.e. corruption of high spatial frequency in the system. Grating based spectrometers also suffer from the keystone error which leads to misregistration between channels. This misregistration error varies channel wise and is also dependent on the spatial field of view for each channel. Sophisticated ground processing can make use of these two phenomenon and can increase the spatial resolution for such systems.

In this work, we have modeled the hyperspectral imaging acquisition process similar to the observation model as proposed by (2). They carried out simulations by assuming high resolution blur as Gaussian shape point spread function and a constant sub-pixel shift between channels. Although the assumption of knowing high resolution blur and constant sub-pixel shifts is fine for simulations, they are seldom known in real scenario. Although for remote sensing satellites, the high resolution blur can be estimated from in-orbit data with stellar acquisition by steering the platform, it is not possible for non-agile platforms. Also, the blur on satellite images is very dynamic and is governed by the intervening atmosphere, jitter on the platform and temperature dependent optics effects. Hence a fixed prior model of the blur cannot be assumed. To handle this scenario, we propose to first remove low resolution channel specific blur by estimating it through blind deconvolution (3), and then use image deconvolution (4) for increasing sharpness of the image. Each channel is then subjected to non-local means filter (5) to increase its signal to noise ratio. Secondly, once channel specific blur from each of the channels are removed, a high resolution panchromatic image covering full spectral range of hyperspectral sensor is constructed by removing the detector sampling blur in the image and using the keystone error of the system in super resolution framework. Finally, a high resolution hyperspectral image is constructed using the generated psuedo high resolution panchromatic images and low-high resolution HySIS image using hypersharpening (1). The approach has been applied to recently launched Indian Hyperspectral satellite named HySIS for VNIR payload data covering wavelength range from 400 nm to 900 nm. PSLV-C43 carring HySIS as its major payload, was launched on November 29, 2018 from First Launch Pad (FLP) of Indian spaceport, Satish Dhawan Space Centre, Sriharikota. Major mission parameters for payload and orbit are given in Table 1.

Table 1. HySIS VNIR Instrument Property

| Parameter | Value |
|---|---|
| | Hysis- VNIR |
| Imaging Principle | Pushbroom-Grating |
| Spectral Range | 400-900 nm |
| Spectral Sampling | ∼10 nm |
| Spectral Resolution | 8-10 nm |
| Spectral channels | 60 |
| Spatial Samples | 1000 |
| Radiometric Resolution | 12 bits |
| Ground Sampling Distance | 30 m in nadir |
| Field of View | ±1.4 |
| Orbit Specs Sun-synchronous with | 630 km Polar near circular orbit Step and Stare of 5:1 with local time of imaging of 10 AM |

A resolution increase of ∼1.3 times has been achieved using this approach which commensurates with the derived telescope modulation transfer function. The image has been compared with LISS-3 data which has native resolution of 23m. The features distinguishable in LISS-3 are also found to be distinguishable in super-resolved HySIS image, which confirms the resolution enhancement. The method is generic and can be applied to increase resolution of similar configuration system.

This paper is organized as follows: Section II gives related work in super-resolution found in the literature. Section III describes the super resolution frame work used and the methodolgy followed to generate the inputs along with the new adaptive prior for super resolution. Section IV shows the experimentation and results obtained along with comparison of super resolved image with LISS-3 data to verify the achieved resolution. Finally section V gives the conclusion.

## 2. Related Work

Over the last several decades, various super resolution methods have been reported in the literature. The methods can be divided into fourier based (6), non-uniform interpolation based(7), iterative back projection based(8), projection on convex sets (POCS) based(9), maximum a posteriori (MAP) based(10) and very recently learning based methods(11). Among those methods, MAP based approaches find special place in literature since they allow the image formation formulation to be incorporated in the model along with the prior knowledge by making use of regularization. In general, the MAP based methods can be divided into two categories. First category is based on variation in fidelity term. The Poisson noise model which is generally followed by images acquired by Charged Coupled Devices (CCD), can be approximated by Gaussian noise model. L2 norm based fidelity model which arise from Gaussian noise assumption is the appropriate model to handle this case. Because it penalizes the error in a quadratic manner, its performance is degraded in the presence of outliers such as salt and pepper noise, transmission loss etc. In case of outliers, L1 norms which comes by white Laplacian noise assumption seems to work better as was shown in(12). Also, it is more robust to registration errors. The major disadvantage of L1 norm is that it produces more observation error, is non-differential and can destroy small textures thereby producing overly smooth images leading to loss of information which can be detrimental for remote sensing images. Recently, M-estimators such as Huber functions(13) which can act as both L1 and L2 norm based on decided threshold have been used to replace fixed norms. The second category of MAP estimate is based on variation in prior. Prior information is added to the cost function to get the desired solution. Regularizers based on L1 based Tikhonov (TV)(15) norm, Total Variation (TV)(14) norm, sparsity based L1 norms including Bilateral TV (BTV)(12) have been proposed. Adaptive version of these norms to reduce noise and at the same time enhance the edge information have been recently proposed such as those based on difference curvature for BTV(16) and TV(17).

In operational satellite data processing, super resolution has been used in SPOT-5 high resolution geometrical (HRG)(18) to increase the spatial resolution from 5 meters(m) to 3m. Skybox(19) also uses algorithm to generate high resolution channel from video acqusitions. Both the satellites acquire images from lower earth orbit. Geofen-4(20) also performs super resolution to increase the resolution the images acquired from geostationery orbit. Recently, CX6-02(21) microsatellite's imaging resolution increased from 2.8m to 1.4m from 700km orbit altitude using super resolution techniques. These super-resolution algorithms work on the assumption that the low-resolution images are single channel and are same spectrally.

An approach to enhancing the spatial resolution of the hyperspectral data cube by combining the SR method with the image fusion technique was proposed in (22). To employ the SR algorithm, the intensities of several subpixel-shifted hyperspectral spectral channels are normalized to a specific channel. However, the normalization procedure introduces unwanted radiometric distortion, which leads to image distortion of the restored single-band HR image. To overcome this, (2) proposed an mathematical model for hyperspectral image acquisition and performed simulation for resolution enhancement by assuming that high resolution blur function and sub-pixel shifts between channels is known.

In recent time, deep learning approaches (23), (24), (25), (26) have gained popularity in hyperspectral image super resolution. These approaches also perform single image super resolution taking 3D cube as input. But their performance is highly dependent on the high resolution training data available for training the model. High resolution hyperspectral data is expensive and seldom available and hence these methods dont find significance in operation data satellite processing of data.

Also, these approaches ignore the the fact that exact high resolution blur is seldom known. The blur is not fixed for all bands as it is function of band specific components like optics, optics aberrations, static behaviour of the system like detector sampling and averaging blur and also of the dynamic platform jitter, atmosphere and some onboard temperature related blur mainly due to optics. Keystone artifact in hyperspectral spectrometer refers to spatial mis-registration due to wavelength-dependent magnification across spectral channels (27). The impact of this artifact on acquired data is that the spatial location of features in different channels are shifted and the spectrum is corrupted. This artifact can arise due to two reasons, chromatic aberration and in-plane detector rotation. Thus, although high resolution blur cannot be known exactly, but keystone in the form of spectral misregistration across the spatial field can be estimated from the lab data using specific targets (28). Also, there are approaches to estimate the keystone error arising due to chromatic aberration and detector in-plane rotation from in-orbit data (29). Once, this mapping is established, it is more or less fixed until on-board temperature changes drastically which effects both the effects.

## 3. Observation Model

Our observation model for deriving low resolution HySIS channels from high resolution PAN channel is similar to that defined in (2) for hyperpsectral images and is given by Equation 1

$$Y_k = S_k D M_k B_k X + N_k \tag{1}$$

where subscript $k$ denote the $k^{th}$ HySIS channel, the matrix $Y_k$ is the low resolution HySIS channel of size $N \times M$, X is the high resolution PAN channel of size $sN \times sM$ where $s$ is the scaling factor, $B_k$ is the high resolution point spread function i.e. the blur which high resolution image undergoes before sampling to produce the corresponding channel, $M_k$ is the motion matrix containing misregistration information between different channels, $D$ is the decimation matrix and $S_k$ is the spectral coefficient matrix which maps the high resolution pseudo PAN channel to low resolution HySIS channels of specific radiometry and $N_k$ is the noise matrix corresponding to low resolution HySIS channel. Given $n$ such low resolution HySIS channels of differing radiometry, the high resolution PAN channel $X$ can be derived by minimizing Equation 2

$$argmin \sum_{k=1}^{n} ||Y_k - S_k D M_k B_k X||_p^p + \phi(X) \tag{2}$$

where the first part of the equation is the likelihood term with norm $p$ and $\phi(X)$ is the prior probability of the desired solution in the MAP framework. $p = 2$ corresponds to Gaussian noise model assumption, $p = 1$ corresponds to Laplacian noise model assumption and $p < 1$ correspond to hyper-laplacian ones. As discussed in the last section, the prior model $\phi(X)$ can take the form of $L2$ norm, $L1$ norm, BTV norm or $Lp$ norm where $p < 1$ (31) for enhancing sparsity.

Having defined the generic observation model, following subsection discusses how values for each of the above terms are being derived.

### 3.1. *Modeling High Resolution Blur*

The blur $B_k$ in a typical sensor is given by Equation 3 as defined in (30)

$$B_k = B_{atmos} * B_{jit} * B_{mot} * B_{optics} * B_{detFoot} * B_{detSamp} \tag{3}$$

where $B_{atmos}$ is the blur to the atmosphere, $B_{jit}$ is due to the jitter on the platform during the image time, $B_{mot}$ is the motion blur due to relative motion between satellite and object between images, $B_{optics}$ is the blur arising due to diffraction limit of the optics, the optical distorions, alignment related blur etc, $B_{detFoot}$ is the blur due to the footprint of the detector and $B_{detSamp}$ is the detector sampling related blur.

Since the blur related to atmosphere and platform jitter are highly dynamic, the overall blur $B_k$ is scene dependent. Also, blur related to optics is channel dependent which further complicates things. To handle these dynamic effects, we calculate channel wise point spread functions using blind deconvolution and correct for this blur with image deconvolution at lower resolution before generating the high resolution image. After correcting for channel specific blur, we use model for only detector sampling blur in the super resolution framework to generate a higher resolution image. Figure 1 shows the image of channel 30 before and after restoration. Thus in the super resolution model mentioned in 1, $B_k$ will only take the value of $B_{detSamp}$, which is the blur due to change in detector sampling. Since we are trying to increase the resolution by a factor of 2, we use Rect function (30) with a support of 2 pixels as model for detector sampling.

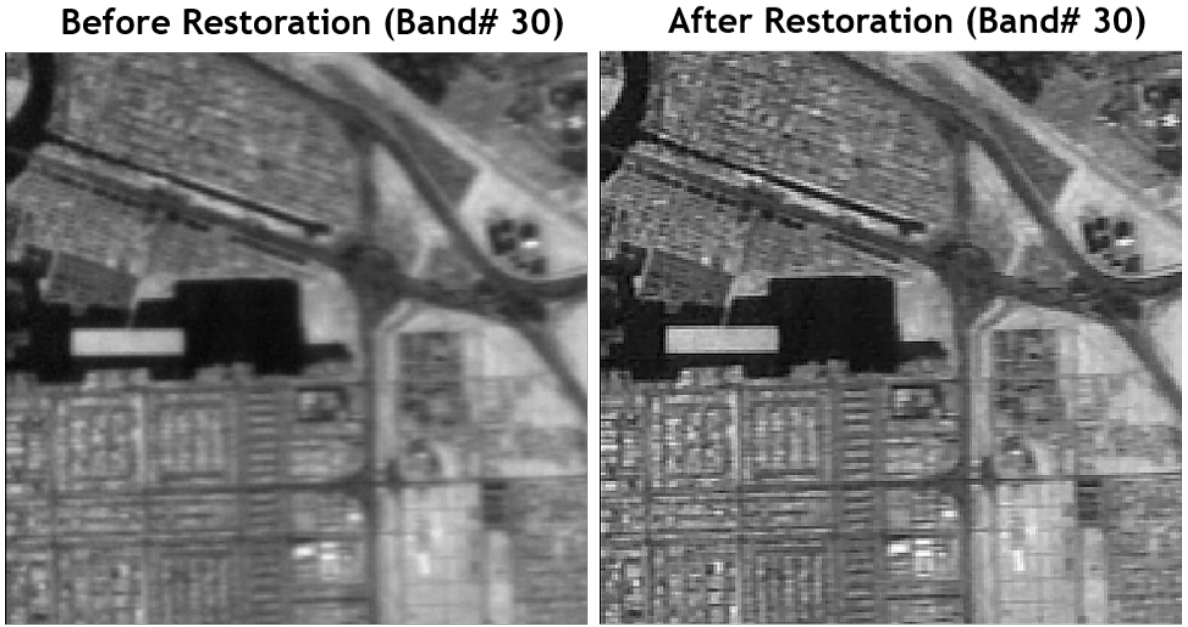

**Figure 1.** Before and After Restoration

### 3.2. *Spectral Coefficient Matrix ($S_k$):*

The spectral coefficient for a certain channel is computed by registering all other channels with respect to this channel and then calculating coefficients as defined in (2), which is given in Equation 4.

$$S_k(i) = I_k(i)/(I_k(i) + \sum_{j \neq k, j=1}^{n} I'_j(i)) \qquad (4)$$

where $I_k(i)$ is the channel for which the spectral coefficient is being estimated, and $I'_j(i)$ are other channels which have been registered to this channel. The spectral coefficients are computed after converting the original digital number(DN) obtained from instrument to radiance value as not doing so and using DN value will result in artifacts in the final image since DN is property of the instrument and is not a physical quantity.

### 3.3. *Prior Model ($\phi(X)$):*

We use BTV operator as the prior model as it preserves edges in our algorithm and also has fewer smoothing effects. The BTV operator is given by Equation 5

$$BTV(X) = \sum_{l=-P}^{P} \sum_{m=0, l+m>0}^{P} \alpha^{|m|+|l|} ||X - S_h^l S_v^m X|| \quad (5)$$

where $S_h^l$ and $S_v^m$ are a translated form of $X$ by $l$ and $m$ pixels in the horizontal and vertical directions respectively. Weight factor $\alpha$ provides spatially decaying effect to regularization. To better preserve edges and textures which are very critical for remote sensing images, we propose to use adaptive weighted BTV prior based on Rmap as given by Equation 6. Rmap was first proposed in (32) to preserve edges and remove smooth regions for kernel estimation. A large value of $r$ indicates that edge with large magnitude exists and small value means that it is relatively smooth region. Figure 2 shows Rmap for one such image patch. The colormaps for original image and Rmap after scaling to 0-1 has also been shown. Rmap shows value close to 1 when sharp transitions in the form of edges are present and value close to 0 in the presence of smooth regions. This prior has been used for edge-preserving filter in (33) but this is the first time that this prior has been deployed in the context of super resolution.

$$w(i) = \exp(-||r(i)||^{0.8}) \quad (6)$$

where $r(i)$ is the Rmap value for the corresponding pixel.

$$r(i) = \frac{||\sum_j \nabla I(i)||_2^2}{\sum_j ||\nabla I(i)||_2^2 + 0.5} \quad (7)$$

where $j$ is a loop which runs in the neighbour of pixel $i$.

### 3.4. *Likelihood Term :*

Because we use robust system level model to estimate registration error and correct for any salt and pepper noise or transmission loss before super resolution, we don't make use of L1 norm but rather use L2 norm as the likelihood term as it is optimal and preserves the textures better. Experiments show that L2 norm performs better than L1 with respect to the sharpness as shown in section IV.
Equation 8 describes the complete minimization function for the whole framework for super resolving multispectral images.

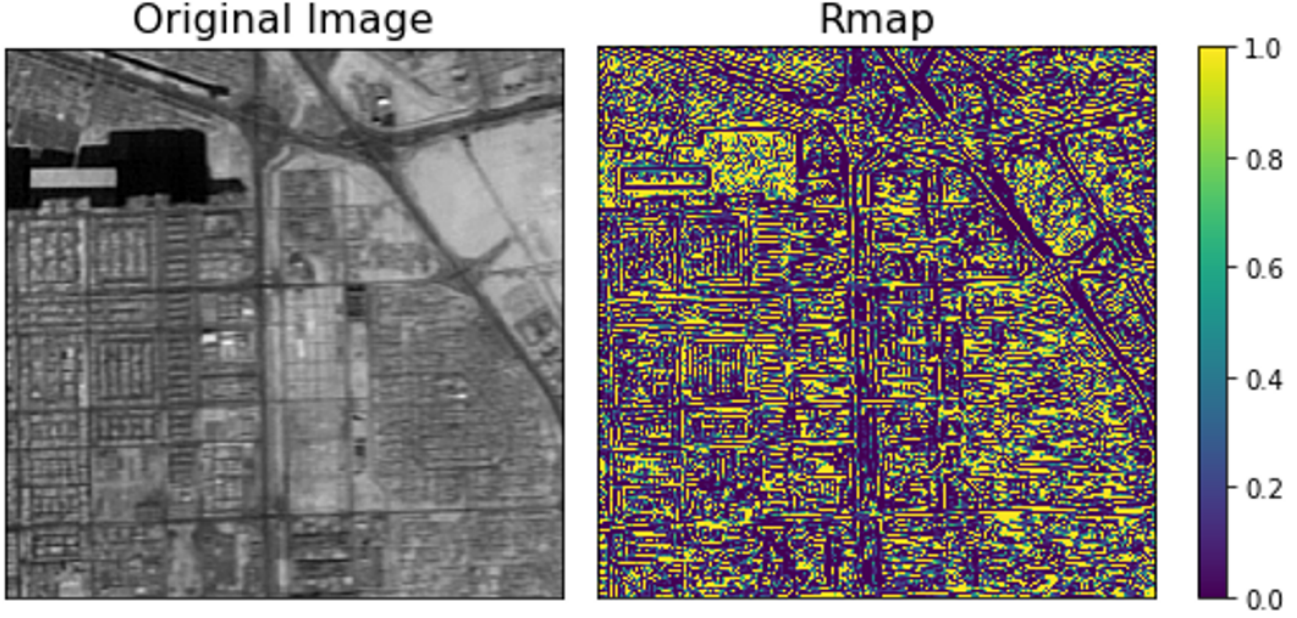

**Figure 2.** Adaptive Prior Model Based on Rmap

$$\hat{X} = argmin \sum_{k=1}^{n} ||Y_k - S_k DM_k B_k X||_2^2 + \lambda w \sum_{l=-P}^{P} \sum_{m=0, l+m>0}^{P} \alpha^{|m|+|l|} ||X - S_h^l S_v^m X|| \quad (8)$$

where $\lambda$ is the parameter to control the strength of regularization parameter as compared to likelihood term.

The above equation is solved using iterative steepest gradient algorithm. Differentiating the above equation with respect to $X$ and putting the derivative to 0 gives the following iterative solution.

$$\hat{X_{n+1}} = \hat{X} - \beta \Big\{ \sum_{k=1}^{n} M_k^T B_k^T D_k^T S_k^T (Y_k - S_k DM_k B_k X) + \lambda w \sum_{l=-P}^{P} \sum_{m=0, l+m>0}^{P} \alpha^{|m|+|l|} (I - S_h^{-l} S_v^{-m}) sign(\hat{X} - S_h^l S_v^m \hat{X}) \Big\} \quad (9)$$

where $\beta$ is the learning rate which is changed adaptively for faster convergence by following scheme: If the cost of the current iteration is lesser than previous iteration, the learning rate is increased by 5%, if more than the last iteration, the learning rate is decreased by 5% and if it is within 1% with respect to previous one for 3 iterations, then the convergence is reached, loop is broken and the solution is returned. The generated high resolution PAN channel is fused with lower resolution channels with smoothing-filter-based intensity modulation(SFIM) fusion technique to generate high resolution HySIS channels.

## 4. Experiment and Results

In all experiments, the learning rate $\beta$ has been set as 0.8, the regularization parameter $\lambda$ has been set to 0.015, gradient weight $\alpha$ term for BTV has been set to 0.2 and $(l, m)$ to (4,4), with maximum number of iterations being 30. These parameters have been determined by experimentation, but are found to perform optimally for all cases. We have taken scenes over Dubai, UAE and Radhanpur, India images during initial phase operations of HySIS. Each of the scenes are 3000 pixels long and 1000 pixels wide.

Modulation Transfer Function and sub-pixel shifts are the two governing factors for the final spatially resolved image or the so called limit of superresolution. Figure 5 show the Modulation Transfer Function of HySIS observed in along and across directions as a function of spatial frequency for different channels. From the graph it is clear that relatively high MTF i.e. ∼11 % is observed at spatial frequency of 0.5 cycles/pixel (i.e. 30 GSD) which indicates the presence of aliased component in the imagery which corrupts the high frequencies. The system MTF goes to very low value at the frequency of 0.7 cycles/pixel, indicating the capability of the system to catch information till those frequencies.

The keystone error was evaluated in-orbit by deriving sub-pixel shifts across the FOV for all spectral bands with respect to channel 30 (central spectral channel) in VNIR instrument. Figure 6 shows the pixel vs pixel error and pixel vs scan error for different bands with respect to band 30 of HySIS. In a typical grating based system, existence of pixel vs pixel error for different bands indicates the combined effect arising due to presence of in-plane detector rotation and keystone in the instrument. Pixel vs scan error across the spectral channels was also observed in HySIS thus giving enough phase shifted information to perform super resolution.

Using the aliasing information of high frequencies as indicated in Figure 5 and the sampling shifts in scan and pixel direction as shown in Figure 6, the native resolution of HySIS is improved.

The performance of the super-resolution approach in presence of proposed prior has been compared to the prominent ones by taking L1 or L2 norm as likelihood term and TV, BTV & our Rmap BTV (RBTV) as prior model. Figure 7 shows the visual results of the experiment for three different patches cropped from the processed image. It can be clearly seen that L1+TV, L1+BTV produce very blurry results. The super resolution performance increase with L2 norm as likelihood term. The best result among the non-adaptive methods is produced by L2+BTV which is able to produce relatively sharper, less noisier and images with less artifacts. Our proposed method L2+RBTV produces the best visual image as it preserves small textures and also decreases noise in the produced output as is evident from the Figure 7. For performing statistical comparison between these methods and to show how the proposed algorithm has helped in resolution and content enhancement, we calculate the two-dimensional fourier transform of the image and then calculate the radial profile of the power spectum. Figure 8 shows the power spectrum thus derived for all frequencies. Power for frequencies 0.25 cycles/pixel to 0.5 cycles/pixels have also been zoomed and shown. L2+RBTV has highest power than other methods at low to high frequencies followed by L2+BTV, which shows the content and resolution enhancement thus achieved further supporting the visual claims. Boost in the both low and high frequency of the image indicates that the overall information of the image has increased thereby qualifying the method.

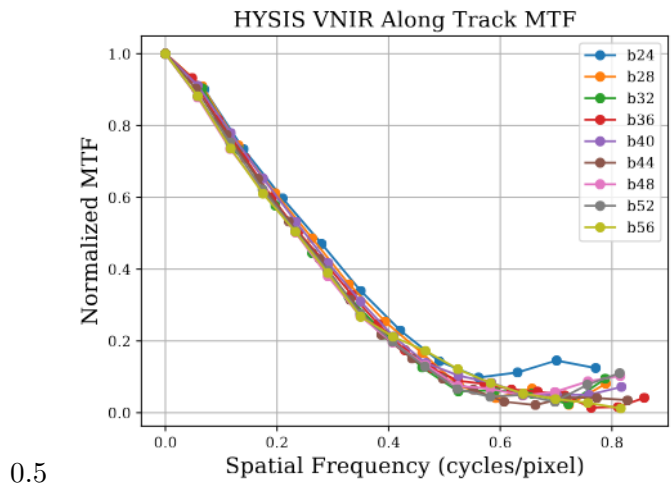

**Figure 3.** Along Track MTFs

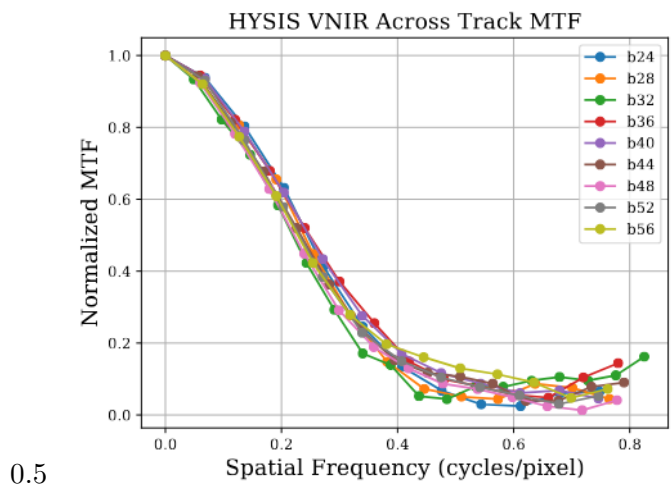

**Figure 4.** Across Track MTFs

**Figure 5.** HySIS VNIR Modulation Transfer Functions

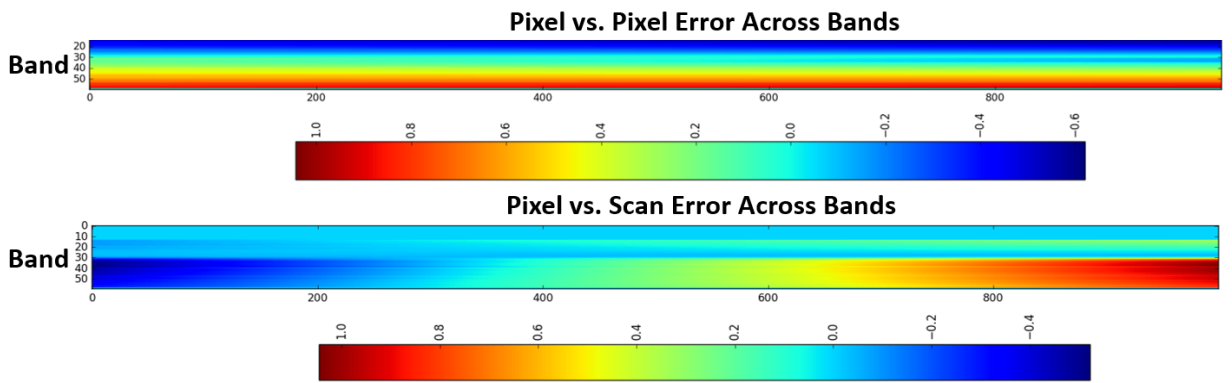

**Figure 6.** VNIR Estimated Keystone Error from In-orbit Data

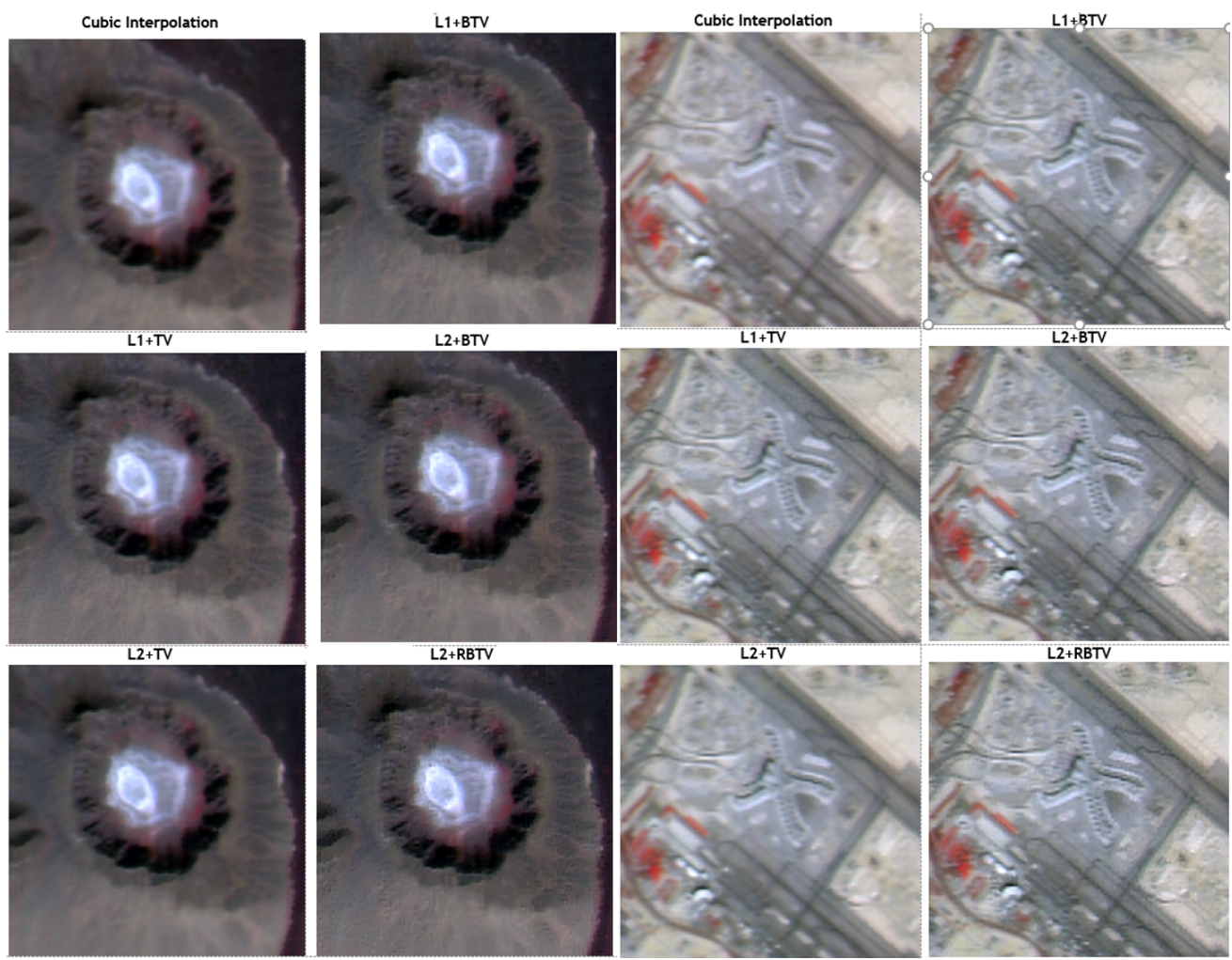

**Figure 7.** Comparison between different methods

In any hyperspectral data, spectral information of the feature is very important. Comparison of spectral features for major features was done before and after super resolution. Figure 9 shows that our super resolution approach does not distort the spectral information. The spectral features of sand and vegetation are very similar.

To validate the method with regards to the achieved resolution, we took data from the Radhanpur site situated in India. The site has got salt lakes, are very useful to evaluate the discriminability capability of the sensor. Figure 10 shows the result before and after super resolution. The salt lakes (marked in red box) which were found to be aliased in the native resolution image have been resolved in the super resolved image indicating that the resolution has increased. To get the measure of actual detectability of the super resolved image , LISS-3 sensor image of Resourcesat 2A was taken of the same area from the same season. The salt lakes which are discriminable in 23m LISS-3 image are also super resolved HySIS image, although the features in LISS-3 are sharper. This is because it has high MTF of  15 % at 23 m sampling where as our super resolved image has very low MTF  1 % at 23 m sampling. But from resolvability perspective, both the images are similar.

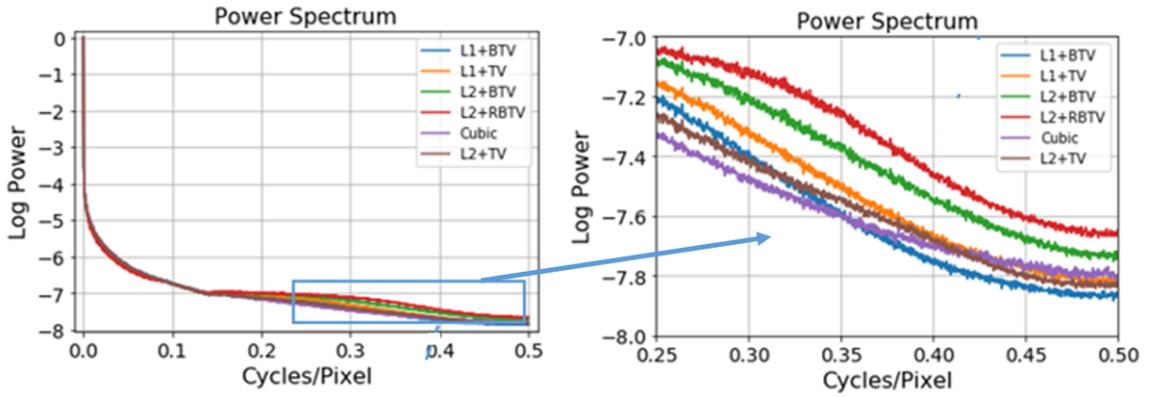

**Figure 8.** Power Spectrum

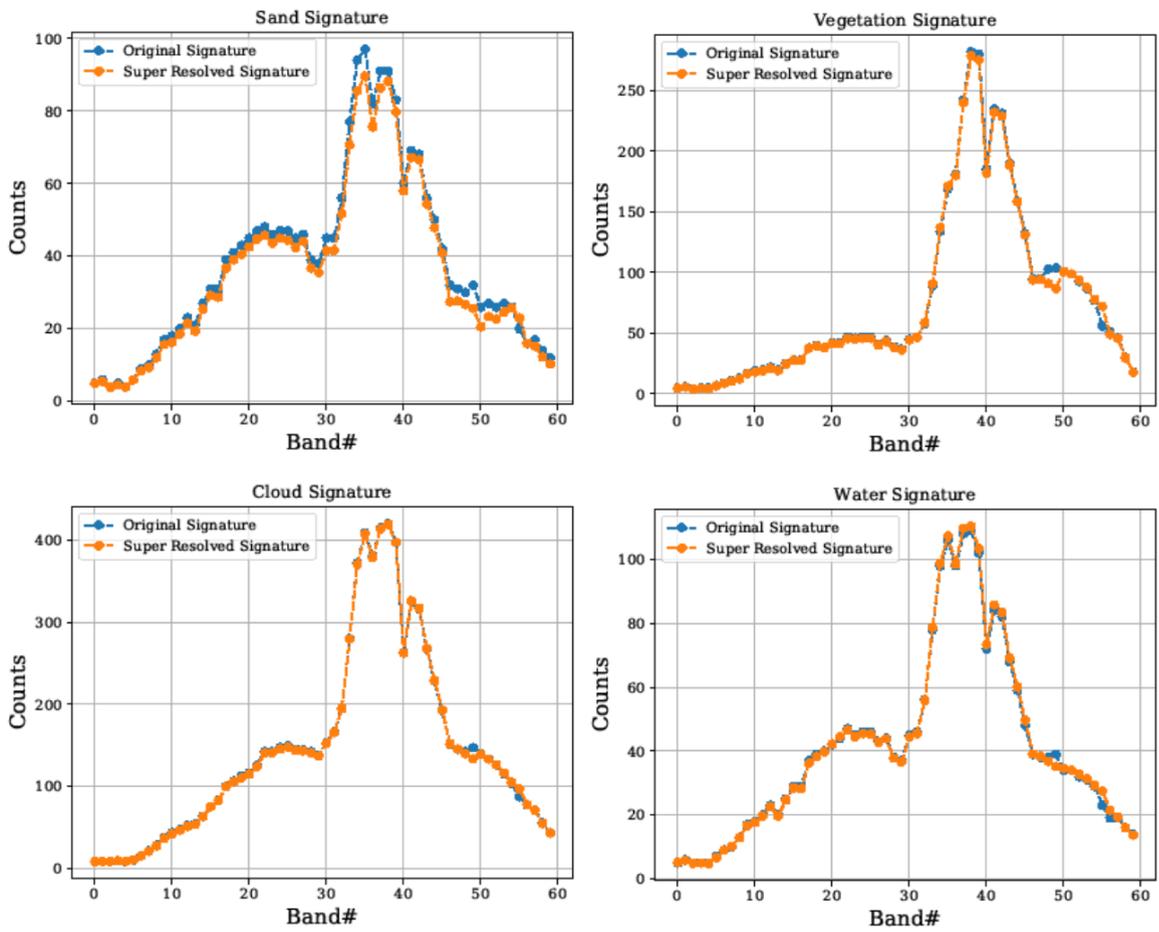

**Figure 9.** Combined Signatures

## 5. Conclusion

This paper proposes an approach to increase the spatial resolution of hyperspectral data without making use of any auxiliary input. For grating based hyperspectral sen-sors, misregistration between spectral channels occur due to instrument intrinsic keystone error which leads to sub-pixel shifts between spectral channels that vary across the spatial field of view. But the high resolution blur for such systems is seldom known due to the dynamic nature of the intervening atmosphere and platform jitter. Also, the optics blur for each channel differs due to differing wavelengths. To handle this dynamic effect, we first estimate the point spread function from lower resolution image using blind deconvolution and correct it using image deconvolution. Then, to handle the remaining sampling related blur, we assume model for it and use it in super resolution framework. The paper also proposes a new adaptive prior and shows how it performs better than those already present in literature. The algorithm is used to increase the resolution of VNIR spectrometer of HySIS, and the results show that it is able to remove the aliasing and lead to a resolution boost of ∼1.3 times (from 30m to 23m) with respect to the input. The algorithm is generic and can be used to increase the resolution of similar configuration systems.

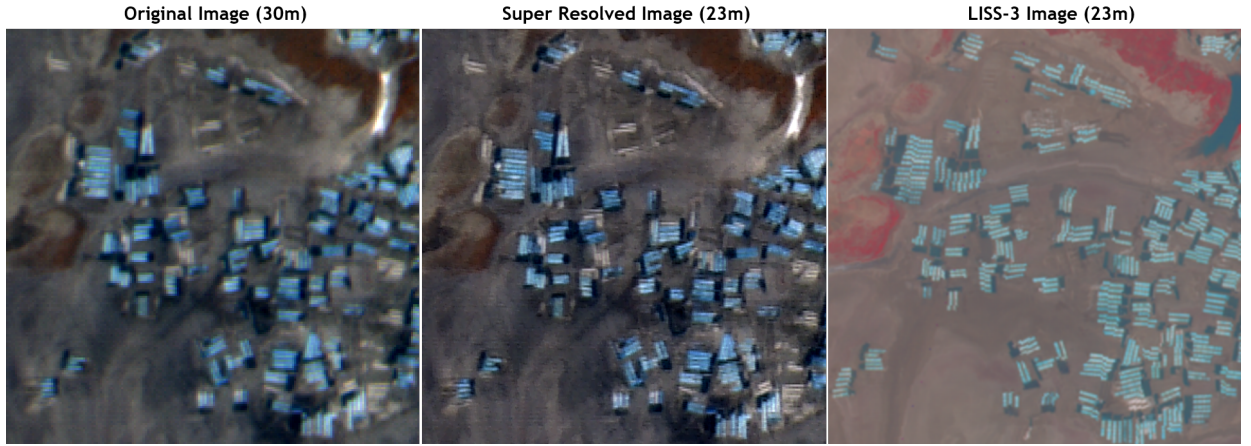

**Figure 10.** Comparison with LISS-3 23 m imagery

## 6. Acknowledgment

The authors would like to acknowledge the Director, Space Applications Centre, Indian Space Research Organisation (ISRO) for providing useful guidance and support for carrying out this work. We also extend our gratitude to the members of HySIS Payload, Mission and Ground Segment Team for their assistance and support.